\documentclass[journal]{IEEEtran}

\usepackage{cite}
\usepackage{graphicx}
\usepackage{amsmath}
\usepackage{amssymb}
\usepackage{array}
\usepackage{dcolumn}
\usepackage[utf8]{inputenc}
\usepackage[T1]{fontenc}

\begin{document}

\title{High efficiency holmium-doped triple-clad fiber laser at 2120~nm}

\author{Bastien Beaumont,
        Pierre Bourdon,
        Alexandre Barnini,
        Louanne Kervella,
        Thierry Robin,
        and~Julien Le Gouët
\thanks{Bastien Beaumont is with the Office National d'Etudes et de Recherches Aérospatiales (ONERA) in Palaiseau, France and with the iXblue Photonics Division in Lannion, France}
\thanks{Pierre Bourdon and Julien Le Gouët are with the Office National d'Etudes et de Recherches Aérospatiales (ONERA) in Palaiseau, France}
\thanks{Alexandre Barnini, Louanne Kervella and Thierry Robin are with iXblue Photonics in Lannion, France}
\thanks{e-mail: {julien.le\_gouet@onera.fr}.}
}

\markboth{Journal of Lightwave Technology}%
{High efficiency holmium-doped triple-clad fiber laser at 2120~nm}

\maketitle

\begin{abstract}
We developed a new holmium-doped triple-clad fiber (Ho-3CF), reducing the clad diameter to improve the overlap between pump and doped-core, and the holmium concentration to reduce the influence of ion clustering on the laser efficiency. We illustrate here the performance of this fiber in a laser oscillator configuration. The laser emission is centered at 2.12~µm by a fiber Bragg grating, and the active fiber is pumped in the clad by a 1.94~µm thulium-doped fiber laser. The slope efficiency reaches 73~\% with respect to absorbed power (60\% considering coupled power), for a maximum signal power of 62~W delivered on a quasi single spatial mode (M$^2$=1.2). To the best of our knowledge, this slope efficiency is the highest reported for a high power clad-pumped fiber laser emitting at a wavelength higher than 2.1~µm. We also analyze experimentally the impact of the pump Numerical Aperture (NA), at the input of the Ho-3CF, on the laser efficiency. Finally, we use our numerical simulation to comment on the best choice of pump wavelength.
\end{abstract}


%
\IEEEpeerreviewmaketitle

\section{Introduction}

\IEEEPARstart{H}{igh} power lasers at eye-safe wavelengths are required in many applications like defence systems, LIDAR, or free-space optical telecommunications. Considering a continuous regime of several kilowatts, the laser systems must be transportable, eye-safe for the operators, and able to concentrate all the power in a specific spot. Those system specifications are equivalent to require a compact laser, robust to vibrations and temperature variations, with high electro-optical efficiency (typically>20\%), emitting at a wavelength higher than 1.4~µm (for atmospheric transmission and eye-safety) in a single spatial mode (generally gaussian), and low heat generation. In addition narrow linewidth amplification must be possible, to allow coherent beam combining of several laser sources.

To date, the highest laser power reported with all these criteria is on the order of the kW, and has been obtained with thulium (Tm) doped fibers \cite{Ehrenreich_2010, Anderson_2021}. However for laser emission beyond 2.1~µm, the power of this type of fiber source might be limited by the thermal load caused by the exothermic cross-relaxation pumping process \cite{Jackson_2004}. Another approach, similar to the ubiquitous erbium:ytterbium codoped fibers, consists in adding Tm sensitizing ions in an holmium (Ho) doped fiber \cite{Allain_91, Motard_2021}. The downside of this approach lies mainly in the difficulty of optimizing simultaneously the two dopant concentrations, and in managing the thermal load.

Another promising approach consists in pumping an Ho-doped fiber close to the absorption peak of the laser transition with several Tm-doped fiber lasers, which are pumped themselves by multimode fiber coupled laser diodes. Since the Tm-based pump lasers can be realized in single-mode fibers, it is possible to couple one such laser into the core of a Ho-doped fiber, allowing an excellent overlap between pump radiation and gain medium. Impressive slope efficiencies of more than 80\% (with respect to absorbed pump power) were reported for core-pumped Ho-doped fiber lasers \cite{Hemming_2016, Baker_2017, Kamradek_SPIE_2020}. Yet for the objective of a kW-class all-fiber laser, coupling multiple high power pump sources into the multimode clad of the active fiber is probably the only option. 

In the clad-pumped configuration, an optical efficiency of about 40\% (with respect to coupled pump power) has already been demonstrated with a record signal power of 407~W in a monolithic architecture \cite{Hemming_CLEO_2013}. The efficiency reached about 50\% in the lower power regime ($<200$~W). In these experiments, the holmium-doped triple-clad fiber had a silica clad diameter of 118~µm \cite{Simakov_2013}. We demonstrated a similar efficiency, reaching a laser signal of 90~W with a fiber with two-fold larger clad diameter, corresponding to a four-fold lower overlap between pump beam and active core \cite{JLG_2020}. This result illustrated the high optical quality of the glass, with low holmium clustering and hydroxyl (OH$^-$) concentration, which are the main contributions to the propagation losses in this type of fiber \cite{Humbach_1996,Wang_2019}.

In this paper, we present the laser performance of our new Ho-doped triple-clad fiber, with a reduced clad diameter of 105~µm, obtained by an improved fabrication process. In the first part, we detail the fiber properties and the experiment setup. In the second part, we present our results with this new fiber and comment on our pump coupling method. In the last part, we discuss about the choice of pump wavelength for the high power applications.

\section{Fiber fabrication and laser setup}
\label{sec_fabrication_setup}

\begin{figure}[ht!]
\centering\includegraphics[width=.7\linewidth]{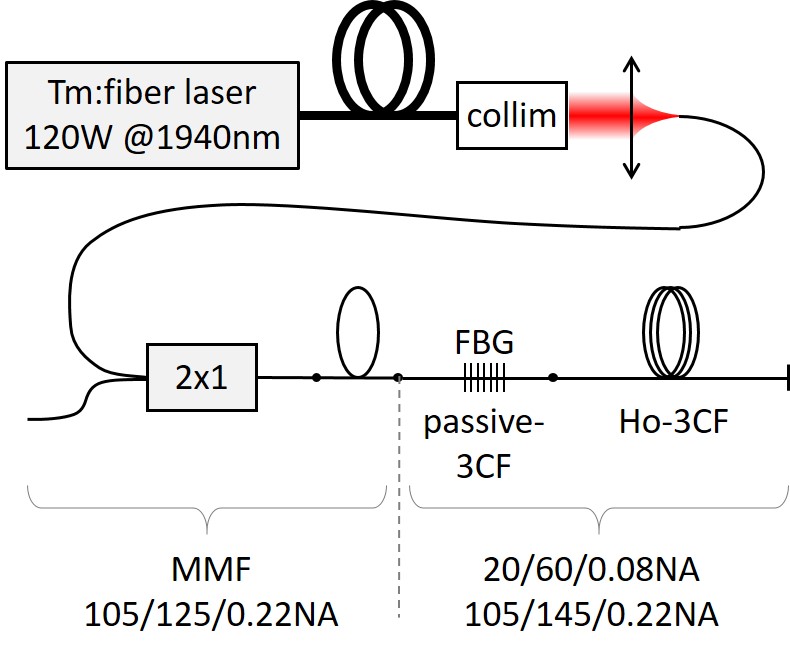}
\caption{Setup of the clad-pumped holmium-doped fiber laser, relying on a single-mode thulium-doped fiber laser pump at 1940~nm.}
\label{fig_dispositif}
\end{figure}

The setup of the clad-pumped Ho-doped fiber oscillator is presented on Figure \ref{fig_dispositif}. The pump beam is provided by a continuous-wave Tm-doped fiber laser from IPG Photonics (TLR-200) emitting at 1.94~µm with 120~W maximum power on a gaussian spatial mode. The output of the fiber laser is an integrated collimator.

We couple the pump beam into a standard multimode silica fiber (MMF) with core/clad diameters 105/125~µm and numerical aperture NA=0.22, using an AR coated dry-silica lens with focal length 100~mm. Each MMF is an input port of a 2x1 combiner, which exits on an identical output MMF.

The holmium-doped fiber has a triple-clad structure, very similar to the fiber described in \cite{JLG_2020}. Its main novelty lies in the new geometry, with a reduced diameter of the silica clad, now at 105~µm flat-to-flat instead of 250~µm and an outer diameter of 145~µm for the  fluorinated clad. The pumping clad diameter now fits the core diameter and NA of the standard MMF used in the combiner. Compared to our previous fiber, the surface ratio between active core and pump beam is thus increased by a factor of almost 6. An interesting point of comparison is the all-fiber laser from DSTG, which produced more than 400~W at 2120~nm. The fiber there had core and clad diameters of 18~µm and 112~µm respectively \cite{Hemming_CLEO_2013}. The doping/pump surface ratio in our fiber is 40\% higher. Another reference to be mentioned is the fiber reported in \cite{Simakov_2017} with core and clad diameters of 14 and 65~µm respectively: the laser reached an efficiency of 67\% with respect to the absorbed pump, which was coupled in free-space and in counter-propagative configuration.

The interest of the reduced clad diameter of our fiber, in addition with the useful fitting with a 105/125/0.22 MMF, is that the pump beam is absorbed on a shorter length, for identical core diameter and holmium concentration $\rho_\text{Ho}$. As long as the main propagation loss are caused by holmium-independent mechanisms (such as OH bond absorption \cite{Humbach_1996}), it is interesting to maintain a high concentration $\rho_\text{Ho}$. However, a high concentration of holmium can lead to the formation of ion clusters, which affect the amplification efficiency through the well-known mechanism of Pair Induced Quenching (PIQ) \cite{Delevaque_1993, Wang_2019}. In a fiber with low propagation loss (e.g. with low OH content), it can become more interesting to reduce the concentration of rare-earth ions, which implies a longer fiber, however compensated by a lower fraction of ions in clusters.

Despite our rare-earth doping process by aqueous solution, we manage to reduce drastically the concentration of hydroxyl groups OH in the fiber with several drying steps (including proper desiccating gaz), to remove the water from the porous layer. In the previous fiber, we estimated from the clad attenuation that the OH concentration was approximately 0.04~ppm \cite{JLG_2020}, and we reach such low values for all rare earth based fibers. Since we did not modify the fabrication process for the core, we assume that the hydroxyl concentration is of the same order of magnitude in our new fiber, i.e. much lower than the 1~ppm reported in the Nufern fiber developed for the DSTG demonstrations \cite{Hemming_CLEO_2013, Simakov_2013}. Therefore we decided to reduce significantly the holmium concentration in the core, in order to reduce accordingly the fraction of ions in pairs. On the other hand, we did not modify notably the aluminum concentration, to preserve the solvation of the remaining holmium ions. 

\begin{figure}[ht!]
\centering\includegraphics[width=.9\linewidth,trim={-1cm 0 0 0},clip]{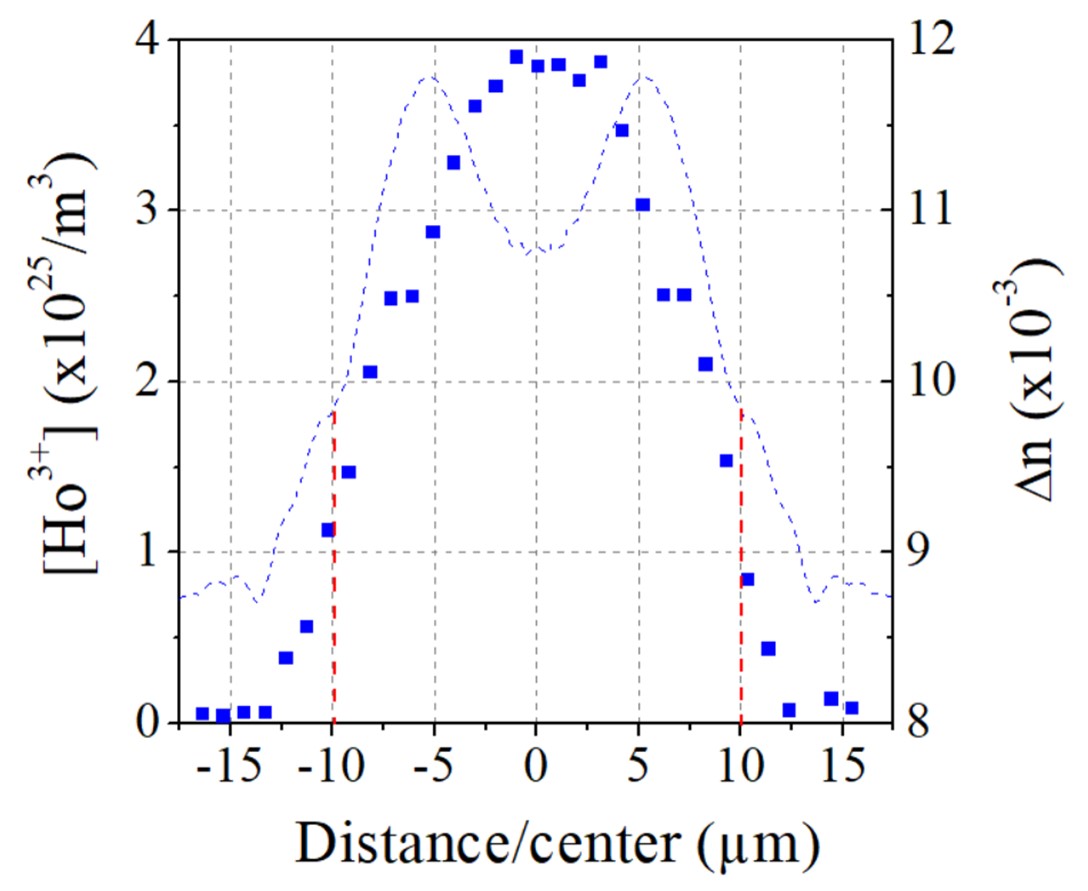}
\caption{Profile of the holmium concentration across the fiber core, obtained by electron micro-probe and an EDS detector. The dotted line represents the refractive index profile in the core (see Fig. \ref{fig_M2_RIP}). The red lines correspond approximately to the guiding core of the fiber.}
\label{fig_compo_A2775}
\end{figure}

By cut-back measurement in the Ho-3CF core at $\lambda_\text{peak}=1950$~nm, we find an absorption $\alpha_\text{Ho}(\lambda_\text{peak})=43 \pm 1$~dB/m for the fundamental optical mode. In the meantime, we measured the chemical composition across the core with an electron micro-probe and energy or wavelength dispersive spectroscopy (EDS and WDS), depending on the atoms. As shown on Figure \ref{fig_compo_A2775}, the average holmium concentration $\rho_\text{Ho}$ in the fiber core is close to $3~10^{25}$~/m$^3$. On a single-clad Ho-doped fiber with similar core composition, we measured a fraction of clustered ions $2k=15 \pm 1\%$ for a holmium concentration of about $4~10^{25}$~/m$^3$ \cite{JLG_2020}. Assuming that the fraction $2k$ is proportional to $\rho_\text{Ho}$ in this range, we estimate that $2k \simeq 11\%$ in our new Ho-3CF. As a comparison, the linear absorption was close to 70~dB/m at 1950~nm in the Ho-doped fiber used in the famous work from DSTG \cite{Hemming_CLEO_2013}, and the fraction of clustered ions was later estimated by numerical simulation to $2k \simeq 30\%$ \cite{Wang_2018}.

As for the peak absorption cross-section for the $^5I_8-^5I_7$ transition in our alumino-silica host, we measure it in the fiber core from the usual formula:
$$ \sigma_\text{abs}(\lambda_\text{peak})=\frac{\alpha_\text{Ho}(\lambda_\text{peak})}{\rho_\text{Ho}.\Gamma(\lambda_\text{peak})}$$
where $\Gamma(\lambda_\text{peak})$ is the overlap between the optical mode guided in the core and the active region. The usual Marcuse formula yields a mode diameter $2w_\text{mode}=23$~µm. Given the active core diameter $D_\text{core}=20$~µm, we thus find an overlap mode/doping $\Gamma(\lambda_\text{peak})=80\%$. The resulting value is $\sigma_\text{abs}(\lambda_\text{peak})=4.1~10^{-25} \text{m}^2$, in good agreement with our previous measurement in a similar host \cite{JLG_2020}. 

The MMF of the combiner output is spliced to a 5~m passive 3CF, coiled with a diameter $D_c\simeq 8$~cm to guarantee an optimal filling of the numerical aperture of the following Ho-3CF. The passive 3CF has essentially the same dimensions as the holmium doped fiber, except that it has no pedestal and that the F-doped silica layer is not octagonal. 

After this pump-mixing fiber, we splice the fiber Bragg grating (FBG) that provides the spectral selectivity at the laser wavelength. The FBG is realized by imprinting a phase mask in the core of a sample of passive 3CF. It exhibits a high reflectivity (HR) of 99\% at 2.12~µm, with a spectral linewidth of about 1~nm. We tested the FBG in two versions, with and without high power packaging, and measured the same results with both samples. In our setup, where the FBG is simply in contact with the optical table, we did not detect any heating at this point of the fiber when injecting the maximal pump power of 110~W in the fiber laser. 

Finally, the low reflectivity of the output coupler of the laser oscillator cavity is obtained either by the Fresnel reflection of a straight cleave on the Ho-doped fiber (R=3\%), or by another FBG imprinted on another sample of passive 3CF. At the fiber laser output, we collimate the signal and residual pump beams, and separate them with a dichroic mirror (HR at 2.12~µm, AR at 1.94~µm).

\section{Laser efficiency}
\label{sec_laser_eff}

\begin{figure}[ht!]
\centering\includegraphics[width=1.15\linewidth,trim={.8cm 0 0 0},clip]{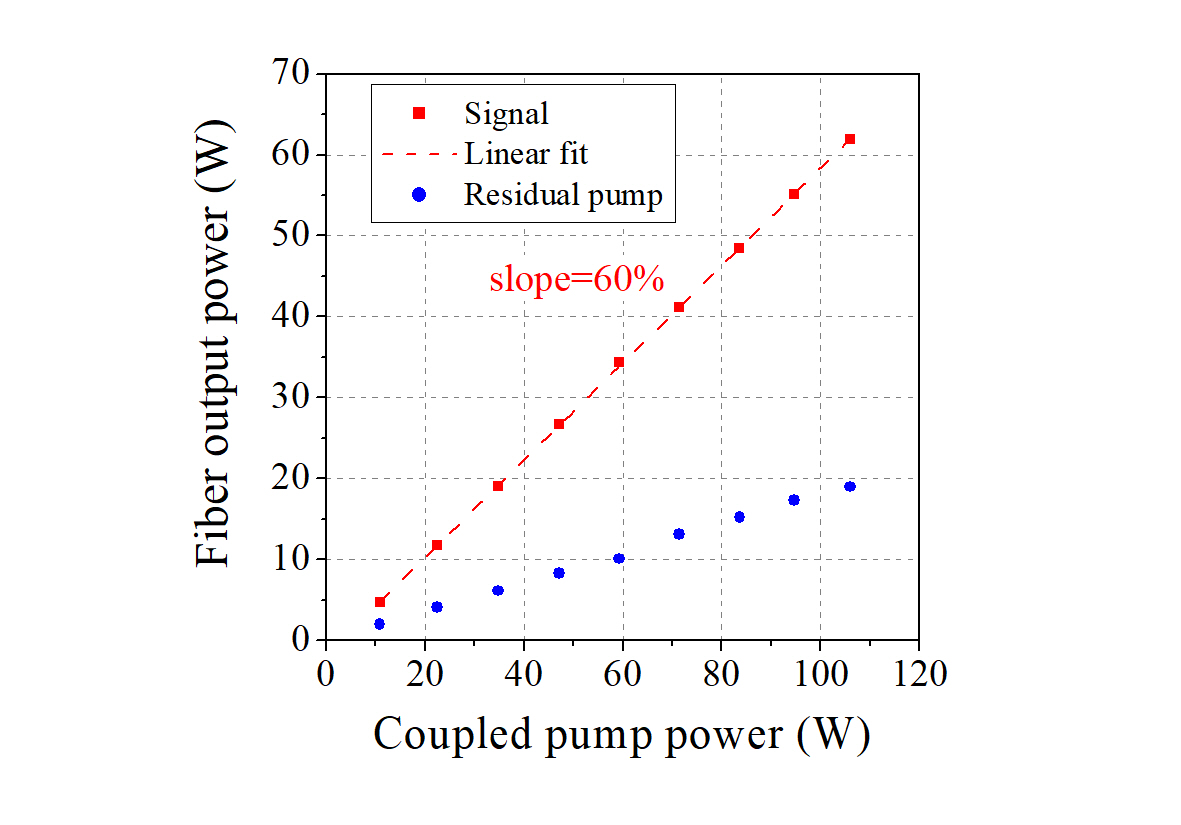}
\caption{Output powers for the signal (red dots) and residual pump (blue), for the optimal fiber length of 9~m, as a function of the coupled pump power. The dashed line is a linear fit with a slope of 60\%.}
\label{fig_out_sig_pump}
\end{figure}

In order to estimate the Ho-3CF optimal length in terms of laser optical efficiency, we use our numerical simulation. Considering approximate values in the simulation for holmium concentration, interaction cross-sections, cluster rate, attenuation (OH groups, silica), and FBG specifications, we find an optimum length close to 8~m. We then precise the Ho-3CF optimal length for laser efficiency by cutback method. Experimentally, we reach an optical-optical efficiency of 60\% for a 9~m fiber, and obtain 62~W at 2120~nm for 106~W pump power (see Figure \ref{fig_out_sig_pump}). The residual pump power values can be also used to calculate the absorbed pump, from which we find that the slope efficiency with respect to the absorbed pump power is 73\%. 

The laser efficiency is thus improved compared to the previous state of the art for clad-pumped all-fiber lasers at 2120~nm \cite{Hemming_CLEO_2013,JLG_2020}. As mentioned above, this performance results from two notable achievements in the fiber fabrication process. First, the ability to realize a complex structure (core/pedestal/silica/fluorinated silica) in a reduced diameter, which allows a better overlap between pump beam and active region compared to our previous work. Second, the low OH concentration in the fiber allows the reduction of the holmium concentration, to reduce the effect of PIQ while maintaining the background attenuation low enough to preserve the laser efficiency. 

Interestingly, the initial measurements were exhibiting a higher efficiency for a slightly shorter optimal length of Ho-3CF, across the same range of pump power. In these experiments, the pump beam was coupled into a quite short length (about 3~m) of standard MMF (105/125/0.22) that was spliced to the passive 3CF containing the FBG. In this configuration, the laser efficiency was close to 65\% for an optimal length of about 8~m. 

When preparing the experiment to couple a second pump laser, we replaced the MMF by the 1~m long input port of the bundle presented above (Fig. \ref{fig_dispositif}). However, with this modified coupling setup the efficiency decreased to about 62\%. We then realized that the pump beam divergence in this new setup was higher than in the previous one, revealing that the 3~m MMF was not sufficient to couple the pump beam in all the spatial modes. This effect is proper to this setup, since our high power pump beams are delivered in free-space from commercial lasers, and must be coupled into MMF to avoid any coupling loss. By using a longer passive MMF, and reducing the coiling diameter, we finally raised the pump beam NA to the value of the clad NA. The laser efficiency of 60\% corresponds to this configuration.

It should be noted that the pump beam in the clad of the Ho-3CF is neither gaussian nor flat-top, so the concept of an absolute beam diameter that would define the numerical aperture is not relevant. In order to determine the NA values at the fiber input, we thus measure the second moment diameter $D_{4\sigma}$ of the far field image of the beam, and assume that the value NA=0.22 corresponds to the maximum value of $D_{4\sigma}$ that can be delivered.

\section{Refractive index profile, beam quality and perspectives}
\label{sec_RIP_M2}

\begin{figure}[ht!]
\centering\includegraphics[width=.9\linewidth]{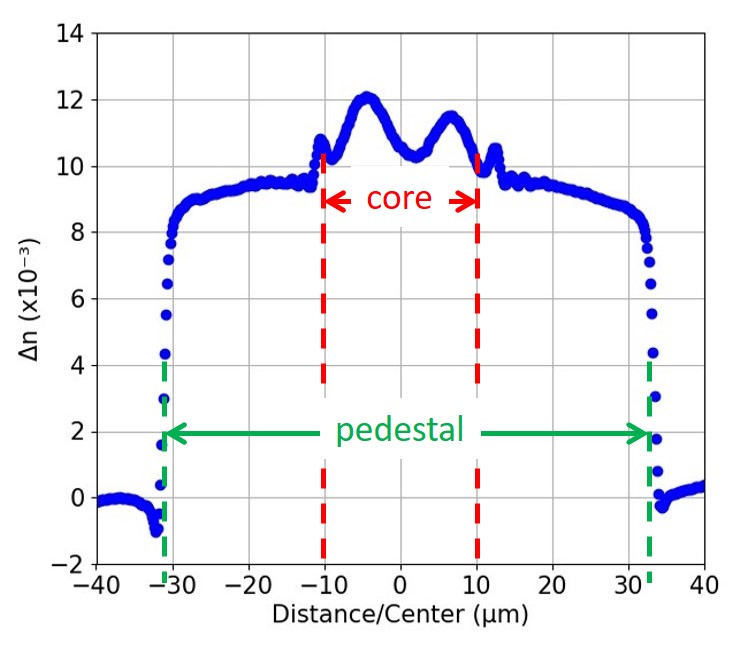}
\caption{Absolute variation of the refractive index profile (RIP) of the holmium-doped alumino-silicate fiber across the pedestal (65~µm diameter) and core (20~µm diameter), compared to the pure silica clad refractive index. The RIP is measured on the fiber side by an optical interferometer (IFA-100, Interfiber Analysis).}
\label{fig_M2_RIP}
\end{figure}

We also checked the beam quality of the output signal, as the measurement of the refractive index across the Ho-3CF diameter revealed a shape quite distinct from the ideal flat step profile (see Figure \ref{fig_M2_RIP}). The beam parameter measurement yielded a surprising good value of M$^2=1.2\pm 0.1$ for the two orthogonal directions. It should be noted however that the M$^2$ value could be notably degraded (up to 1.9 in one of the directions) by tightening slightly the Ho-3CF. This sensitivity of the beam quality to the fiber coiling or position may be due to the uneven index profile of the fiber core.

Flattening the refractive index profile and avoiding the pedestal are some of the perspectives of improvement for our holmium-doped fibers. We are presently looking for alternative host materials for the fiber core, and one of the promising candidates is alumino-phospho-silicate, where the silica glass is codoped with almost identical concentrations of phosphorous and aluminum oxides. The introduction of AlPO$_4$ units in silica is famous for allowing the refractive index reduction in the rare-earth-doped core \cite{Digiovanni_1989}, so the pedestal could be suppressed. Moreover this material reduces notably the clustering of rare-earth ions and allows higher laser efficiency, as already demonstrated with other rare-earths such as erbium \cite{Likhachev_2009, Jebali_2014}, ytterbium \cite{Liu_2018}, or neodymium \cite{Barnini_2020}. Finally, the multiphonon absorption in the fiber core may also be reduced in the AlPO$_4$ codoped silica \cite{Lord_2020}. This would be a determining advantage of this host, as it is becoming the new limiting contribution of propagation loss in the 2~µm domain.

\section{On the pump wavelength}
\label{sec_discussion}

In several recent communications \cite{Tench_SPIE_2021,Tench_JLT_2021,Tench_ECOC_2021}, the choice of a pump wavelength at higher energy (typically around 1850~nm) has been presented as an innovative advantageous approach, as compared to the 1950~nm wavelength used in lasers dedicated to high power \cite{Hemming_CLEO_2013, Friebele_2014, Pal_2016, Holmen_2021, JLG_2020}. Here we propose a brief analysis of the interests and drawbacks of these wavelength, based on numerical simulations.

For rare-earth ions, infrared optical transitions are available between electronic energy states that contain themselves several Stark shifted substates. As it is well-known, pumping towards the highest energy states can convert an apparent two-level system into a quasi three-level system \cite{Desurvire_EDFA, Paschotta_1997}. A higher population inversion and better gain can thus be obtained by pumping at wavelengths shorter than the peak absorption $\lambda_\text{peak}$, where the probability of emission is also at its maximum. This is exactly the case for a fiber short enough to avoid pump depletion, and where the population inversion remains almost constant.

Let us consider now the spectral variation of the saturation intensity $I_\text{sat}=h\nu/\left[\tau.(\sigma_\text{abs}+\sigma_\text{em})\right]$, where $\tau$ is the excited state lifetime and $\sigma_\text{abs}$ and $\sigma_\text{em}$ are the absorption and emission cross-sections \cite{Saleh_1990}. As a consequence of the highest absorption and emission at $\lambda_\text{peak}$ (about 976~nm for Yb, 1530~nm for Er, 1950~nm for Ho), $I_\text{sat}$ reaches its lowest value for this wavelength. For example, $I_\text{sat}$ is higher by a factor of 6 at 920~nm compared to 976~nm in an Yb-doped fiber, and a factor of almost 10 at 1850~nm compared to 1950~nm in Ho-doped fiber. Therefore, for a high pump intensity $I_\text{pump}$ (compared to $I_\text{sat}$ at any wavelength), the pump beam will bleach the fiber on a longer section if it is set at $\lambda_\text{peak}$, and thus provide population inversion on a longer section, and finally higher gain integrated over the fiber. 

\begin{figure}[ht!]
\centering\includegraphics[width=.9\linewidth]{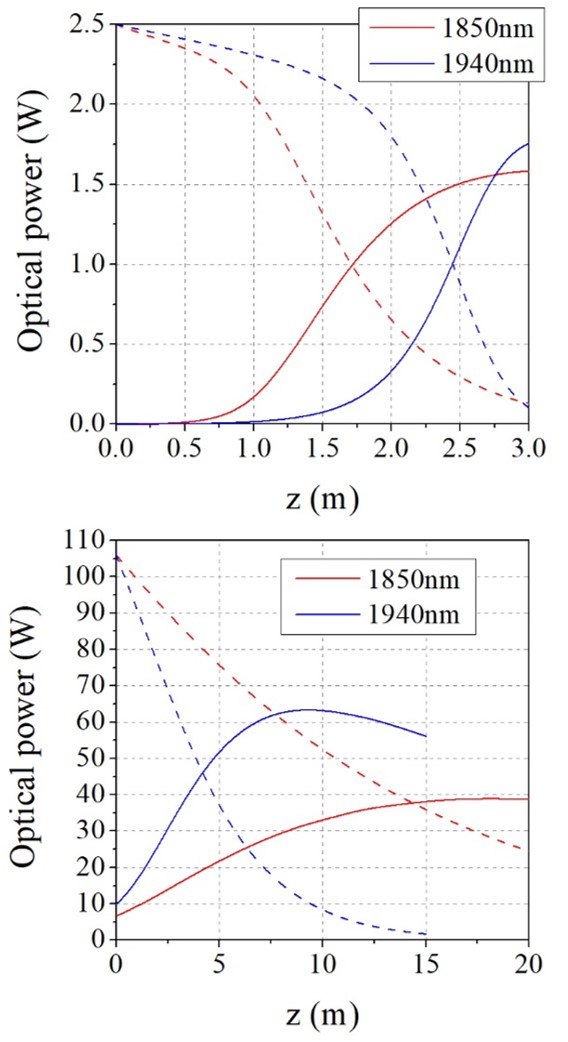}
\caption{Numerical simulation of the laser signal at 2120~nm (straight lines) and pump powers (dotted lines), for a pump beam at 1850~nm (red) or at 1940~nm (blue). Top: amplifier described in \cite{Tench_JLT_2021} for a hypothetical Ho$^{3+}$-doped fiber without ion clustering (2k=0). Bottom: fiber laser oscillator in the configuration reported above.}
\label{fig_simus}
\end{figure}

We illustrate this aspect by two examples: first the core-pumped fiber amplifier described in \cite{Tench_JLT_2021}, but without ion clustering, and second the clad-pumped high power laser oscillator reported above (taking clusters into account). On Figure \ref{fig_simus} we present the corresponding results of our numerical simulation for the signal and pump powers (straight and dotted lines respectively), for a pump beam at 1850 or 1940~nm (red and blue lines). 

Although hypothetical, the case of the holmium-doped fiber without ion clustering is interesting, as it brings to a conclusion that differs from the cited works \cite{Tench_ECOC_2021, Tench_SPIE_2021, Tench_JLT_2021}. As detailed above, a given pump power saturates more efficiently the holmium ions at 1940~nm (close to $\lambda_\text{peak}$) than at 1850~nm. Even though the maximum population inversion is lower for 1940~nm pumping, it is provided on a longer path, which eventually results in a higher output signal. For the parameters mentioned in \cite{Tench_JLT_2021}, but taking into account the Ho ion pairs, our simulation confirms the results presented in the cited article.

Finally, the example of clad-pumped high power laser reminds that shifting the pump wavelength out of the absorption peak requires to increase the fiber length. In particular, since the Ho$^{3+}$ absorption is about 5 times lower at 1850~nm than at $\lambda_\text{peak}$, the optimal fiber laser would be almost 5 times longer at 1850~nm in absence of any fiber loss. This has several negative consequences: first, in copropagative pumping, the population inversion is high at the very beginning of the fiber, and becomes depleted long before the output, so there is less gain available towards the fiber end. In a real fiber laser, various propagation losses (multi-phonon transitions, OH$^-$absorption, pair-induced quenching...) contribute to reducing the optimal length and the maximum power. Therefore, the shorter the fiber and the less affected the output power, which comes as an advantage again of pumping at 1940~nm instead of 1850~nm. In addition, in the context of narrow-linewidth high power laser amplifiers, nonlinear effects such as Stimulated Brillouin Scattering are more limiting for longer fiber length. The same arguments are also valid for Yb or Er-doped fibers for example, which explains why most high power fiber sources are pumped at the absorption peak of the laser transition \cite{Yu_2016, Hawkins_2021, Jebali_2014}. As a conclusion, defining a systematic judicious choice of pump wavelength seems difficult, as it depends on the specific configuration of the laser.


\section{Conclusion}
By modifying the fabrication process of our holmium-doped fiber, we managed to reduce the diameter of the silica clad down to 105~µm, instead of 250~µm for the previous successful realization. Thanks to a better overlap between pump and active core, a lower holmium concentration and non-saturable absorption by PIQ, we achieve a signal power of 62~W at 2120~nm, with a slope efficiency of 73\% with respect to the absorbed pump power at 1940~nm (60\% considering the coupled pump power). To our knowledge, this is the highest value reported so far for a clad-pumped single-mode continuous-wave holmium-doped fiber laser. We also report on the influence of the pump numerical aperture at the fiber input on the laser efficiency. Finally, we use our numerical simulation to show why the pump wavelength should be set to the peak holmium absorption for a high-power clad-pumped fiber laser.

\appendices

\section*{Acknowledgment}

The authors are thankful to the Direction Générale de l'Armement for financial support.

\ifCLASSOPTIONcaptionsoff
  \newpage
\fi

\bibliographystyle{IEEEtran}
\bibliography{IEEEabrv,bib_2022_LaserHo_VERSOplus}

\begin{thebibliography}{10}
\providecommand{\url}[1]{#1}
\csname url@samestyle\endcsname
\providecommand{\newblock}{\relax}
\providecommand{\bibinfo}[2]{#2}
\providecommand{\BIBentrySTDinterwordspacing}{\spaceskip=0pt\relax}
\providecommand{\BIBentryALTinterwordstretchfactor}{4}
\providecommand{\BIBentryALTinterwordspacing}{\spaceskip=\fontdimen2\font plus
\BIBentryALTinterwordstretchfactor\fontdimen3\font minus
  \fontdimen4\font\relax}
\providecommand{\BIBforeignlanguage}[2]{{%
\expandafter\ifx\csname l@#1\endcsname\relax
\typeout{** WARNING: IEEEtran.bst: No hyphenation pattern has been}%
\typeout{** loaded for the language `#1'. Using the pattern for}%
\typeout{** the default language instead.}%
\else
\language=\csname l@#1\endcsname
\fi
#2}}
\providecommand{\BIBdecl}{\relax}
\BIBdecl

\bibitem{Ehrenreich_2010}
T.~Ehrenreich, R.~Leveille, I.~Majid, K.~Tankala, G.~Rines, and P.~F. Moulton,
  ``1k{W}, all-glass {Tm}:fiber laser,'' in \emph{proc. SPIE Photonics West},
  2010.

\bibitem{Anderson_2021}
B.~M. Anderson, J.~Soloman, and A.~Flores, ``1.1 {kW}, beam combinable thulium
  doped all-fiber amplifier,'' in \emph{Fiber {Lasers} {XVIII}: {Technology}
  and {Systems}}, M.~N. Zervas and C.~Jauregui-Misas, Eds.\hskip 1em plus 0.5em
  minus 0.4em\relax SPIE, Mar. 2021, p.~7.

\bibitem{Jackson_2004}
S.~D. Jackson, ``Cross relaxation and energy transfer upconversion processes
  relevant to the functioning of 2 µm {Tm3}+-doped silica fibre lasers,''
  \emph{Optics Communications}, vol. 230, no. 1-3, pp. 197--203, Jan. 2004.

\bibitem{Allain_91}
J.~Allain, M.~Monerie, and H.~Poignant, ``High-efficiency {CW}
  thulium-sensitised holmium-doped fluoride fibre laser operating at 2.04
  µm,'' \emph{Electronics Letters}, vol.~27, no.~17, p. 1513, 1991.

\bibitem{Motard_2021}
A.~Motard, C.~Louot, T.~Robin, B.~Cadier, I.~Manek-Hönninger, N.~Dalloz, and
  A.~Hildenbrand-Dhollande, ``Diffraction limited 195-{W} continuous wave laser
  emission at 2.09~µm from a {Tm} $^{\textrm{3+}}$ , {Ho} $^{\textrm{3+}}$
  -codoped single-oscillator monolithic fiber laser,'' \emph{Optics Express},
  vol.~29, no.~5, p. 6599, 2021.

\bibitem{Hemming_2016}
A.~Hemming, N.~Simakov, M.~Oermann, A.~Carter, and J.~Haub, ``Record efficiency
  of a holmium-doped silica fibre laser,'' in \emph{Conference on {Lasers} and
  {Electro}-{Optics}}.\hskip 1em plus 0.5em minus 0.4em\relax San Jose,
  California: OSA, 2016, p. SM3Q.5.

\bibitem{Baker_2017}
\BIBentryALTinterwordspacing
C.~C. Baker, E.~J. Friebele, A.~A. Burdett, D.~L. Rhonehouse, J.~Fontana,
  W.~Kim, S.~R. Bowman, L.~B. Shaw, J.~Sanghera, J.~Zhang, R.~Pattnaik,
  M.~Dubinskii, J.~Ballato, C.~Kucera, A.~Vargas, A.~Hemming, N.~Simakov, and
  J.~Haub, ``Nanoparticle doping for high power fiber lasers at eye-safer
  wavelengths,'' \emph{Opt. Express}, vol.~25, no.~12, pp. 13\,903--13\,915,
  2017. [Online]. Available:
  \url{http://www.opticsexpress.org/abstract.cfm?URI=oe-25-12-13903}
\BIBentrySTDinterwordspacing

\bibitem{Kamradek_SPIE_2020}
M.~Kamrádek, J.~Aubrecht, M.~Jelínek, M.~Frank, P.~Peterka, P.~Honzátko,
  J.~Mrázek, P.~Vařák, O.~Podrazký, F.~Todorov, V.~Kubeček, and
  I.~Kašík, ``Holmium-doped fibers for efficient fiber lasers at 2100 nm,''
  in \emph{{OSA} {High}-brightness {Sources} and {Light}-driven {Interactions}
  {Congress} 2020 ({EUVXRAY}, {HILAS}, {MICS})}.\hskip 1em plus 0.5em minus
  0.4em\relax OSA, 2020, p. MTh3C.5.

\bibitem{Hemming_CLEO_2013}
A.~Hemming, N.~Simakov, A.~Davidson, S.~Bennetts, M.~Hughes, N.~Carmody,
  P.~Davies, L.~Corena, D.~Stepanov, J.~Haub, R.~Swain, and A.~Carter, ``A
  monolithic cladding pumped holmium-doped fibre laser,'' in \emph{CLEO:
  2013}.\hskip 1em plus 0.5em minus 0.4em\relax Optical Society of America,
  2013, p. CW1M.1.

\bibitem{Simakov_2013}
N.~Simakov, A.~Hemming, W.~A. Clarkson, J.~Haub, and A.~Carter, ``A
  cladding-pumped, tunable holmium doped fiber laser,'' \emph{Optics Express},
  vol.~21, no.~23, pp. 28\,415--28\,422, 2013.

\bibitem{JLG_2020}
J.~Le~Gouët, F.~Gustave, P.~Bourdon, T.~Robin, A.~Laurent, and B.~Cadier,
  ``Realization and simulation of high-power holmium doped fiber lasers for
  long-range transmission,'' \emph{Optics Express}, p.~14, 2020.

\bibitem{Humbach_1996}
O.~Humbach, H.~Fabian, U.~Grzesik, U.~Haken, and W.~Heitmann, ``Analysis of
  {OH} absorption bands in synthetic silica,'' \emph{Journal of Non-Crystalline
  Solids}, vol. 203, pp. 19 -- 26, 1996, optical and Electrical Propertias of
  Glasses.

\bibitem{Wang_2019}
J.~Wang, N.~Bae, S.~B. Lee, and K.~Lee, ``Effects of ion clustering and excited
  state absorption on the performance of ho-doped fiber lasers,'' \emph{Opt.
  Express}, vol.~27, no.~10, pp. 14\,283--14\,297, May 2019.

\bibitem{Simakov_2017}
N.~Simakov, ``Development of components and fibres for the power scaling of
  pulsed holmium-doped fibre sources,'' Ph.D. dissertation, University of
  Southampton, 2017.

\bibitem{Delevaque_1993}
E.~{Delevaque}, T.~{Georges}, M.~{Monerie}, P.~{Lamouler}, and J.-F. {Bayon},
  ``Modeling of pair-induced quenching in erbium-doped silicate fibers,''
  \emph{IEEE Photonics Technology Letters}, vol.~5, no.~1, pp. 73--75, 1993.

\bibitem{Wang_2018}
J.~{Wang}, D.~{Yeom}, N.~{Simakov}, A.~{Hemming}, A.~{Carter}, S.~B. {Lee}, and
  K.~{Lee}, ``Numerical modeling of in-band pumped {Ho}-doped silica fiber
  lasers,'' \emph{Journal of Lightwave Technology}, vol.~36, no.~24, pp.
  5863--5880, 2018.

\bibitem{Digiovanni_1989}
D.~DiGiovanni, J.~MacChesney, and T.~Kometani, ``Structure and properties of
  silica containing aluminum and phosphorus near the {AlPO4} join,''
  \emph{Journal of Non-Crystalline Solids}, vol. 113, no.~1, pp. 58--64, 1989.

\bibitem{Likhachev_2009}
M.~E. Likhachev, M.~M. Bubnov, K.~V. Zotov, D.~S. Lipatov, M.~V. Yashkov, and
  A.~N. Guryanov, ``Effect of the {AlPO}\_4 join on the pump-to-signal
  conversion efficiency in heavily {Er}-doped fibers,'' \emph{Optics Letters},
  vol.~34, no.~21, p. 3355, Nov. 2009.

\bibitem{Jebali_2014}
M.~A. Jebali, J.-N. Maran, and S.~LaRochelle, ``264 {W} output power at 1585 nm
  in {Er}–{Yb} codoped fiber laser using in-band pumping,'' \emph{Optics
  Letters}, vol.~39, no.~13, p. 3974, Jul. 2014.

\bibitem{Liu_2018}
S.~Liu, R.~Zhu, J.~Wang, F.~Jing, A.~Lin, K.~Peng, H.~Zhan, L.~Ni, X.~Wang,
  Y.~Wang, Y.~Li, J.~Yu, and L.~Jiang, ``3 {kW} 20/400 {Yb}-{Doped}
  {Aluminophosphosilicate} {Fiber} with high stability,'' \emph{IEEE Photonics
  Journal}, vol.~10, no.~5, pp. 1--8, Oct. 2018.

\bibitem{Barnini_2020}
A.~Barnini, K.~Le~Corre, L.~Kervella, T.~Robin, P.~Guitton, M.~Laroche, and
  S.~Girard, ``Low numerical aperature large-mode-area neodymium-doped fibers
  fabricated by {SPCVD} and {ASD} for laser operation near 920nm,'' in
  \emph{Optical {Components} and {Materials} {XVII}}, M.~J. Digonnet and
  S.~Jiang, Eds.\hskip 1em plus 0.5em minus 0.4em\relax San Francisco, United
  States: SPIE, Mar. 2020, p.~19.

\bibitem{Lord_2020}
M.-P. Lord, L.~Talbot, O.~Boily, T.~Boilard, G.~Gariépy, S.~Grelet,
  P.~Paradis, V.~Boulanger, N.~Grégoire, S.~Morency, Y.~Messaddeq, and
  M.~Bernier, ``Erbium-doped aluminophosphosilicate all-fiber laser operating
  at 1584 nm,'' \emph{Optics Express}, vol.~28, no.~3, p. 3378, 2020.

\bibitem{Tench_SPIE_2021}
R.~E. Tench, J.-M. Delavaux, and C.~Romano, ``Novel highly efficient in-band
  pump wavelengths for {Ho}-doped fiber amplifiers,'' in \emph{Fiber {Lasers}
  {XVIII}: {Technology} and {Systems}}, M.~N. Zervas and C.~Jauregui-Misas,
  Eds.\hskip 1em plus 0.5em minus 0.4em\relax SPIE, 2021, p.~11.

\bibitem{Tench_JLT_2021}
R.~E. Tench, W.~Walasik, and J.-M. Delavaux, ``Novel {Highly} {Efficient}
  {In}-{Band} {Pump} {Wavelengths} for {Medium} {Slope} {Efficiency}
  {Holmium}-{Doped} {Fiber} {Amplifiers},'' \emph{Journal of Lightwave
  Technology}, pp. 1--1, 2021.

\bibitem{Tench_ECOC_2021}
\BIBentryALTinterwordspacing
R.~E. Tench, W.~Walasik, A.~Amavigan, and J.-M. Delavaux, ``Performance
  {Benefits} of 1860 nm vs. 1940 nm {Pumping} of {Holmium}-doped {Fibres} with
  {Significant} {Ion} {Pairing},'' in \emph{2021 {European} {Conference} on
  {Optical} {Communication} ({ECOC})}.\hskip 1em plus 0.5em minus 0.4em\relax
  IEEE, 2021, pp. 1--4. [Online]. Available:
  \url{https://ieeexplore.ieee.org/document/9605871/}
\BIBentrySTDinterwordspacing

\bibitem{Friebele_2014}
E.~J. Friebele, C.~G. Askins, J.~R. Peele, B.~M. Wright, N.~J. Condon,
  S.~O'Connor, C.~G. Brown, and S.~R. Bowman, ``Ho-doped fiber for high energy
  laser applications,'' in \emph{Proc. SPIE}, vol. 8961, 2014, p. 896120.

\bibitem{Pal_2016}
D.~Pal, A.~Dhar, R.~Sen, and A.~Pal, ``All-fiber {Holmium} {Laser} at 2.1~µm
  under in-band {Pumping},'' in \emph{13th {International} {Conference} on
  {Fiber} {Optics} and {Photonics}}.\hskip 1em plus 0.5em minus 0.4em\relax
  Kharagpur: OSA, 2016, p. Tu3E.2.

\bibitem{Holmen_2021}
L.~G. Holmen and H.~Fonnum, ``Holmium-doped fiber amplifier for pumping a
  {ZnGeP} $_{\textrm{2}}$ optical parametric oscillator,'' \emph{Optics
  Express}, vol.~29, no.~6, p. 8477, Mar. 2021.

\bibitem{Desurvire_EDFA}
E.~Desurvire, \emph{Erbium-Doped Fiber Amplifiers: Principles and
  Applications}.\hskip 1em plus 0.5em minus 0.4em\relax Wiley, 1994.

\bibitem{Paschotta_1997}
R.~Paschotta, J.~Nilsson, A.~Tropper, and D.~Hanna, ``Ytterbium-doped fiber
  amplifiers,'' \emph{IEEE Journal of Quantum Electronics}, vol.~33, no.~7, pp.
  1049--1056, Jul. 1997.

\bibitem{Saleh_1990}
A.~Saleh, R.~Jopson, J.~Evankow, and J.~Aspell, ``Modeling of gain in
  erbium-doped fiber amplifiers,'' \emph{IEEE Photonics Technology Letters},
  vol.~2, no.~10, pp. 714--717, Oct. 1990.

\bibitem{Yu_2016}
C.~X. Yu, O.~Shatrovoy, T.~Y. Fan, and T.~F. Taunay, ``Diode-pumped narrow
  linewidth multi-kilowatt metalized {Yb} fiber amplifier,'' \emph{Optics
  Letters}, vol.~41, no.~22, p. 5202, Nov. 2016.

\bibitem{Hawkins_2021}
T.~W. Hawkins, P.~D. Dragic, N.~Yu, A.~Flores, M.~Engholm, and J.~Ballato,
  ``Kilowatt power scaling of an intrinsically low {Brillouin} and thermo-optic
  {Yb}-doped silica fiber,'' \emph{Journal of the Optical Society of America
  B}, vol.~38, no.~12, p. F38, Dec. 2021.

\end{thebibliography}

\end{document}